\documentstyle[aps,multicol,prl,amssymb]{revtex}
\renewcommand{\narrowtext}{\begin{multicols}{2} \global\columnwidth20.5pc}
\renewcommand{\widetext}{\end{multicols} \global\columnwidth42.5pc}

\input{epsf.tex}
\def\inseps#1#2{\def\epsfsize##1##2{#2##1} \centerline{\epsfbox{#1}}}

\def \attn #1 {{\sl $\bullet$ #1 $\bullet$}}
\begin{document}
\draft

\title{Stable Skyrmions in two-component Bose-Einstein
condensates}
\author{Richard A. Battye$^{(1)}$, N. R. Cooper$^{(2)}$ and Paul M. Sutcliffe$^{(3)}$ }
\address{(1) Department of Applied Mathematics and Theoretical
Physics, Centre for Mathematical Sciences, \\ Wilberforce Road, Cambridge
CB3 OWA, United Kingdom.}
\address{(2) Theory of Condensed Matter Group, Cavendish Laboratory, 
Madingley Road, Cambridge CB3 0HE,
United Kingdom.}
\address{(3) Institute of Mathematics, University of Kent at
Canterbury, Canterbury CT2 7NF, United Kingdom.}
\date{September 24, 2001}

\maketitle

\begin{abstract}

We show that stable Skyrmions exist in two-component atomic
Bose-Einstein condensates, in the regime of phase separation. Using
full three-dimensional simulations we find the stable Skyrmions with
topological charges $Q=1$ and 2, and compute their properties. With
reference to these computations we suggest the salient features of an
experimental setup in which they might realized.

\end{abstract}

\pacs{PACS Numbers: 03.75.Fi, 67.57.Fg}

\narrowtext

Experimental advances in the formation and control of ultra-cold
atomic gases are allowing detailed studies of the properties of
Bose-Einstein condensates (BECs) containing multiple components.
Multi-component condensates have been formed by the simultaneous
trapping and cooling of atoms in distinct hyperfine or spin
levels\cite{MyattBGCW97,HallMEWC98,Stamper-KurnACIMSK98,StengerISMCK98},
and there are prospects of condensed mixtures of different bosonic
atomic species\cite{BurkeBEG98,BlochGMHE01}.  The extra internal
degrees of freedom introduced by the multiple components lead to a
much richer phenomenology than in the single component case.

One key feature of multi-component BECs is that, in general, their
condensed groundstates are described by spinor order
parameters\cite{OhmiM98Ho98,Leggett01}.  This raises the possibility that
these systems could support novel topological defects and textures.
The existence of topological vortices\cite{OhmiM98Ho98,LeonhardtV00} and
monopoles\cite{StoofVK2001}, in which the topology is imposed by
boundary conditions, has been noted.  However, no localised
topological textures, whose structure and topology is fixed only by
simple energetic stability, have been identified.  In
Ref.~\cite{RuostekoskiA2001} it was shown that a configuration with
the topology of a ``Skyrmion''~\cite{Skyrme} (a topological soliton of
the $S^3\rightarrow S^3$ map\cite{rajaraman}) can be imprinted in a
two-component Bose-Einstein condensate, using a carefully designed
sequence of Rabi transitions. The subsequent time evolution indicated
that this object is energetically unstable. In related
work\cite{AlKhawajaS01}, it was demonstrated that a
spherically-symmetric Skyrmion in a 2- or 3-component ferromagnetic
BEC is also unstable to collapse.  In
Refs.\cite{RuostekoskiA2001,AlKhawajaS01} a particular choice of
interaction coefficients was considered.  However, the general
multi-component system is described by all mutual two-body scattering
lengths\cite{Leggett01}.  Given the range of parameter space available
it is natural to ask: Are there {\it any} conditions under which
Skyrmions are stable?

In this Letter, we present the results of extensive numerical studies
of a fully 3-dimensional system that show that stable Skyrmions do
exist in two-component BECs\cite{footnote}. We find that stable
Skyrmions exist under the condition that {\it phase separation}
occurs.  We therefore expect Skyrmions to be stable in the
experimentally relevant $^{87}$Rb $|F=2,m_F=1\rangle$ $|1,-1\rangle$
system\cite{HallMEWC98}. As we will show, the Skyrmion can be viewed
as a quantised vortex ring in one component close to whose core is
confined the second component carrying quantised circulation around
the ring.  As such, the configurations closely resemble the ``cosmic
vortons''~\cite{DavisS89} which may have formed in the early universe
from superconducting cosmic strings~\cite{DavisS88I} providing a
further interesting link with between low-temperature laboratory
experiments and cosmology (see, for example, Ref.~\cite{Kibble}).

We consider a two-component system, in which the number of atoms of
each component is separately conserved on the timescale of interest,
and assume that the many-body wavefunction can be written as a simple
condensate\cite{Leggett01}
\begin{equation}
\Psi(\{r_1,r_2\ldots\}) = \prod_{i=1}^{N_1+N_2}\left[\psi_1(\bbox{r}_i)|1\rangle + \psi_2(\bbox{r}_i) |2\rangle\right]\,,
\end{equation}
where the states $|\alpha=1,2\rangle$ are the eigenstates of the two
components (for simplicity of presentation, we take these to be two
hyperfine levels).  We are interested in {\it stable} stationary
states, which are minima of the total energy
\begin{eqnarray}
\nonumber
E  & =  & \int d^3\bbox{r}\left[\sum_\alpha \frac{\hbar^2}{2m}
  |\nabla \psi_\alpha|^2 + V_\alpha(\bbox{r}) |\psi_\alpha|^2 \right. \\ & &  +  \left.\frac{1}{2}\sum_{\alpha,\beta}U_{\alpha\beta}|\psi_\alpha|^2|\psi_\beta|^2
  \right] \,,
\label{eq:ham}
\end{eqnarray}
where the matrix $U_{\alpha\beta}$ is determined by all mutual
$s$-wave scattering amplitudes\cite{Leggett01}.  Since the number of
atoms of each component is conserved, the minimisation is performed
with constraints on
\begin{equation}
N_\alpha = \int d^3 \bbox{r} |\psi_\alpha|^2
\end{equation}
that is, having separate chemical potentials $\mu_1\neq \mu_2$.

The parameters entering (\ref{eq:ham}) allow many different regimes.
We will study solutions confined to a region close to the centre of a
large trap that is loaded with component-$|1\rangle$. We
may then neglect the effects of the confining potentials, and consider
an infinite uniform system with boundary condition
\begin{equation}
\left(\begin{array}{c} \psi_1 \\ \psi_2\end{array}\right) \stackrel{|r|\rightarrow\infty}{\longrightarrow} \left(\begin{array}{c} \sqrt{\rho_0} \\ 0\end{array}\right)\,,\label{eq:bc}
\end{equation}
where $\rho_0$ is the density at the centre of the trap.
Furthermore, we concentrate on situations in which the interactions
are repulsive, and only weakly dependent on species $U_{11} \sim
U_{12} \sim U_{22}$. In this case,  the
stationary solutions vary on scales much larger than the healing
length $\xi\equiv (2m\rho_0 \bar U \hbar^2)^{-1/2}$.  Consequently, we
can neglect variations in the total density $\rho(\bbox{r})$, and write
\begin{equation}
   \left(\begin{array}{c}
\psi_1 \\ \psi_2\end{array}\right) = \sqrt{\rho_0} 
\left(\begin{array}{c} \cos(\theta/2) e^{i\phi_1}\\
\sin(\theta/2)e^{i\phi_2}\end{array}\right) \,,
\label{eq:spinor2}
\end{equation}
in which we choose a parameterisation of the remaining freedom in
terms of $\theta \in [0,\pi]; \phi_\alpha\in[0,2\pi]$.  The field
(\ref{eq:spinor2}) and boundary condition (\ref{eq:bc}) allow a
topological classification of (non-singular) field configurations, in
terms of the winding number for the map $S^3\rightarrow S^3$
\begin{equation}
Q = \frac{1}{8\pi^2} \int \sin\theta 
\nabla_i\theta\nabla_j\phi_1\nabla_k\phi_2\;
\epsilon_{ijk}d^3\bbox{r}\,.
\label{eq:Q}
\end{equation}

We search for topological solitons by minimising, within each
topological subspace $Q$, the energy functional
\begin{eqnarray}
\nonumber
E-E_0
  & =&  \int d^3\bbox{r}  \left\{\frac{\hbar^2\rho_0}{2m}\left[
\frac{1}{4}|\nabla\theta|^2 + \cos^2(\theta/2)|\nabla\phi_1|^2\right.\right.
 \\ & &  \left.  \left. + \sin^2(\theta/2)|\nabla\phi_2|^2\right]
 +\Delta \sin^2\theta \right\}
\,,
\label{eq:fun}
\end{eqnarray}
where 
$\Delta \equiv \frac{1}{8}\rho_0^2\left[2U_{12}-U_{11}-U_{22}\right]$
with a constraint on
\begin{equation}
N_2 = \frac{1}{2}\rho_0 \int d^3\bbox{r} (1-\cos\theta)\,.
\label{eq:num}
\end{equation}
The constraint on $N_1$ is implicit in the choice of
$\rho_0$. In (\ref{eq:fun}) the zero of the energy $E_0$
is a constant which depends on $U_{11}$, $U_{22}$
and the constrained numbers $N_1$ and $N_2.$

There are two important length-scales in the problem
\begin{equation} R_2 \equiv \left(\frac{N_2}{\rho_0}\right)^{1/3}
\hskip0.25cm ;
\hskip0.25cm  \xi_\Delta \equiv
\sqrt{\frac{\hbar^2\rho_0}{2m\Delta}} .
\end{equation}
$R_2$ may be interpreted as the typical size of a cloud of
component-$|2\rangle$; $\xi_\Delta$ may be interpreted as the width of
the transition region separating regions of component $|1\rangle$ and
$|2\rangle$.  Our neglect of the confining potential is justified
provided $\mbox{max}(R_2,\xi_\Delta)$ is small compared to the overall
size of the cloud; the restriction to constant density
(\ref{eq:spinor2}) is justified for $\xi_\Delta \gg \xi$, that is, 
$|\Delta|\ll \bar U$.  The dimensionless ratio
$\eta \equiv ({ R_2}/{\xi_\Delta})$,
which is a measure of the amount of component-$|2\rangle$, is the
parameter that characterises the solutions of (\ref{eq:fun}), with the
energies taking the form
\begin{equation}
E_Q(\Delta, N_2) =
\left(\frac{\hbar^2\rho_0 R_2}{m}\right)
{\cal E}_Q(\eta)\,,
\label{eq:energy}
\end{equation}
where ${\cal E}_Q(\eta)$ is a dimensionless function.

Consider first {\it non}-topological solutions $(Q=0)$.  For the
limit, $\nabla\phi_1=\nabla\phi_2=0$, the problem reduces to a 3D
double sine-Gordon model, which is known to have soliton solutions for
$\Delta>0$, and $\eta>1.92$\cite{kosevich}. For large $N_2$, the
soliton can be viewed as a spherical domain of component-$|2\rangle$
with radius $\sim R_2$, separated from the region of component
$|1\rangle$, by a domain wall of width $\sim\xi_\Delta$. Therefore, it
describes a spherical inclusion of component-$|2\rangle$ arising from
{\it phase-separation}.

\begin{figure}
\inseps{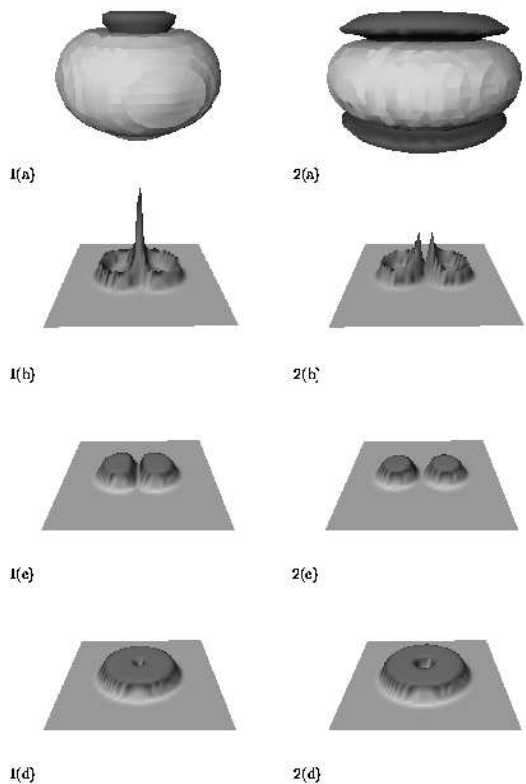}{0.48}
\caption{The results of the 3D numerical relaxation for $Q=1$ (on the
left) and $Q=2$ (on the right): (a) the iso-surfaces of the
topological charge density (dark shading) and the number density of
component-$|2\rangle$ (light shading); (b) the energy density on a
plane containing the axis of symmetry; (c) the number density of
component-$|2\rangle$ on the same plane and (d) the number density of
component-$|2\rangle$, on a plane cutting the axis of
symmetry. All figures correspond to $\eta \approx 9 $. }
\label{fig:3d}
\end{figure}

In contrast, there are no spherically symmetric {\it topological}
solutions ($Q\ge 1$); such solutions having at most axial
symmetry. We have found such solutions for $Q=1$ and $2$ using a full
3D simulation.  Briefly, we evolved the full hyperbolic evolution
equations of the system corresponding to the energy functional
(\ref{eq:fun}) while constraining (\ref{eq:num}) using a Lagrange
multiplier on a discretized grid with $100^3$ points. Starting from a
wide range of different initial conditions without any particular
symmetry, but with the same value of $Q$, the kinetic energy was
periodically removed so as to prevent oscillatory motion and
effectively simulate gradient flow. This technique had previously been
applied successfully to Skyrmions in the original model of
Skyrme~\cite{BS1} and also to the closely related Hopf
solitons~\cite{BS2}. The solutions evolved quickly to the required
value of $N_2$ and the subsequent relaxation was achieved after an
acceptable period of time. As a simple test of the procedure we were
able to reproduce quantitatively the results of Ref.~\cite{kosevich}
for $Q=0$.

The results for $Q=1,2$ are presented in Fig.~\ref{fig:3d} for the
phase separation regime, $\Delta > 0$, illustrating the existence of
stable topological solitons. We have plotted 3D iso-surfaces of the
topological charge density and the number density of species
$|2\rangle$ (the integrands of (\ref{eq:Q}) and (\ref{eq:num})
respectively). Both illustrate the axial symmetry of the solutions:
for this precise value of the threshold the topological charge density
appears to be mostly localized around a central line in the shape of a
``bolt'', although for a different threshold the figure would have
appeared to be a vastly different since there is also some topological
charge associated with the thin ($\sim\xi_{\Delta})$ axially symmetric
domain wall.  The number density is clearly in the form of an axially
symmetric ring encircling the ``bolt'' of topological charge density.

Slices through the energy density (the integrand of (\ref{eq:fun}))
and the number density in the plane perpendicular to the ring
illustrate the precise details of the solutions --- in particular the
two length-scales $R_2$ and $\xi_{\Delta}$. The energy density, whose
morphology is very similar to that of the topological charge density,
is largely localized along the line through the centre of the vortex
ring. However, one can clearly see that, emanating from this line,
there is a shell of energy associated with the domain wall. The
corresponding slice through the number density reaffirms this: it is
almost zero away from the ring and rises sharply to close to one in
the interior. An alternative view of the solution is a slice
perpendicular to the line through the centre of the ring.

In each case, the configuration can be viewed as a (unit) quantised
vortex ring in component-$|1\rangle$, close to whose core is confined
a circulating ring of component-$|2\rangle$ with quantised circulation
(1 or 2 units, for $Q=1,2$). The kinetic energy of the vortex ring and
the surface tension of the domain wall separating the two components
energetically favour shrinking the ring to zero size. This is balanced
by the kinetic energy of the circulating core, allowing for stable
configurations of non-zero radius.  In this respect our solutions have
similar qualitative form to the unstable configurations discussed in
Ref.~\cite{RuostekoskiA2001} for $\Delta=0$. Clearly, it is phase
separation, induced when $\Delta>0$ that allows stabilization of these
solutions.

Our 3D numerical studies indicate that the stable Skyrmions with
$Q=1,2$ are cylindrically symmetric. To determine accurate values for
the energies of these configurations, we have performed simulations
using a cylindrical ansatz consistent with the 3D results (and
consistent with the variational equation for $\phi_2$):
$\theta(\rho,z), \phi_1(\rho,z), \phi_2 = m \chi$ [$(\rho,\chi,z)$ are
cylindrical polar co-ordinates]. When $m\neq 0$, the remaining 2D
problem for $\theta(\rho,z), \phi_1(\rho,z)$ has finite energy
solutions provided $\theta=0$ on $\rho=0$ and at
$\rho^2+z^2\rightarrow\infty$; this allows the configurations to be
characterised by the winding number, $n$, of $\phi_1$ around the
half-plane.
Thus, there is a topological classification of the cylindrical ansatz
in terms of the pair of integers $(m,n)$, which describe the
circulation of the vortex ring in component-$|1\rangle$ ($n$) and the
circulation of component-$|2\rangle$ in the core ($m$).  The winding
number for the $S^3\rightarrow S^3$ map (\ref{eq:Q}) is $Q=nm$.

\begin{figure}
\inseps{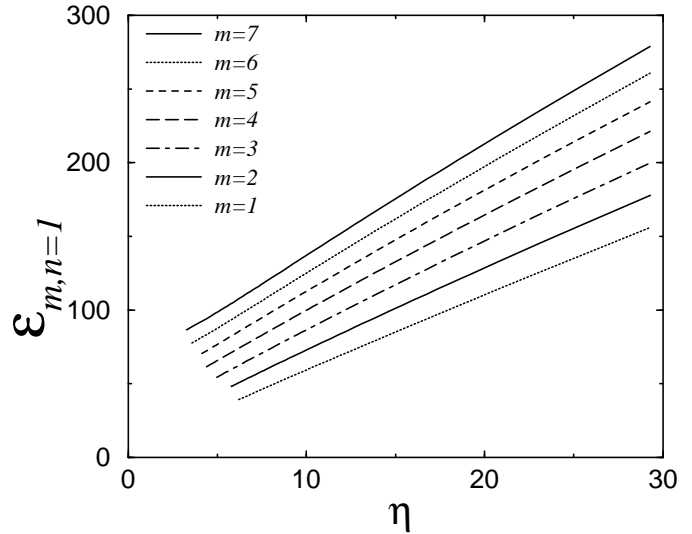}{0.5}
\caption{Scaled energies ${\cal E}_{m,n=1}$ of Skyrmions as a function
of $\eta$. The circulation of the vortex ring is $n=1$; the
circulation of the core is $m=1\ldots 7$ (from bottom to top), such
that the topological charge is $Q=m$. The lines terminate below some
critical value of $\eta$ at which the solitons become
unstable. Results are from calculations using the cylindrical ansatz
on a $100\times 200$ grid with $R_2 = 80$ lattice spacings.}
\label{fig:energies}
\end{figure}

We have determined the soliton energies
from numerical minimisations using a standard conjugant gradient
routine. The results agree, to within numerical accuracy, with those of the
3D simulations. For $m=0$ we recover the 
non-topological soliton.  We find stable topological solitons for {\it
all} $m>0$, but {\it only} for $|n|=1$.  The energies of the
solitons ${\cal E}_{m,n=1}(\eta)$ are plotted in
Fig.~\ref{fig:energies}. For $Q\geq 3$ (up to the largest we
have studied) we find stable topological solitons  within the cylindrical
ansatz. These solutions may not be stable
to non-axisymmetric perturbations which is currently under
investigation.

As described above, the Skyrmions can be viewed as generalisations of
quantised vortex rings (for example, in superfluid $^4$He) to a
two-component condensate.  One distinction we wish to emphasise is
that the stable Skyrmions of Figs.~\ref{fig:3d},~\ref{fig:energies} do
not move in space, whereas conventional vortex rings move at constant
(non-zero) velocity.  We can find Skyrmion configurations that do
propagate, by adding a constraint to fix the impulse
\begin{equation}
P_i  =  \frac{\hbar}{2i} \int d^3\bbox{r} [r_j\nabla_i\psi_\alpha^*\nabla_j\psi_\alpha -
r_j\nabla_j\psi_\alpha^*\nabla_i\psi_\alpha ]
\label{eq:P}
\end{equation}
which is conserved by the time-dependent Gross-Pitaevskii
equation. Under these (dissipationless) dynamics, one can show that the
configuration that minimizes the energy at fixed $\bbox{P}$ translates
in space at a constant velocity $\bbox{v} = \partial E/\partial
\bbox{P}$, where $E$ is its energy. Dimensional analysis allows the
energy of a state with constraints on $Q$, $N_2$ and $\bbox{P}$ to be
written as ${\cal E}_Q(\eta,p)$ in Eq.(\ref{eq:energy}), where the new
dimensionless parameter is $p \equiv |\bbox{P}|/ ( \hbar \rho_0^{1/3}
N_2^{2/3})$.
\begin{figure}
\inseps{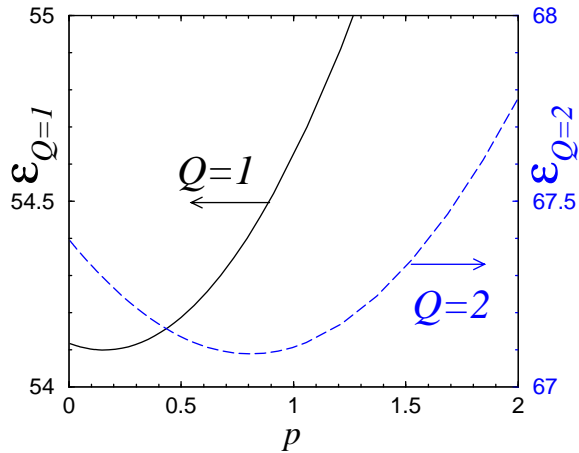}{0.42}
\caption{Scaled energies of the stable Skyrmions with $Q=1$ (solid
line, left axis) and $Q=2$ (dashed line, right axis) as a function of
scaled impulse, $p$, for $\eta = 9$.  The minima correspond to the
stationary Skyrmions. Calculations were performed using the
cylindrical ansatz on a $100\times 200$ grid with $R_2=80$.}
\label{fig:mtm}
\end{figure}
In Fig.~\ref{fig:mtm} we show the dispersion relations of 
Skyrmions with $Q=1$ and $2$.  The stationary configurations of
Figs.~\ref{fig:3d} and~\ref{fig:energies} correspond to the local
minima ($\bbox{v}=\partial E/\partial \bbox{P} =0$).  Away from these
minima, the velocities of the Skyrmions are non-zero. For $p\gg 1$,
the configurations resemble conventional vortex rings in which the
radius of the ring is large compared to the core-size.

In conclusion, we summarise the conditions for observation of stable
Skyrmions.  In the limit that we have studied $\Delta \ll \bar U$, we
find solutions provided (i) $\Delta > 0$, and (ii) $N_2$ is
sufficiently large that $\eta$ is above the termination points of the
curves in Fig.~\ref{fig:energies}.  Under these conditions, we expect
that atomic Bose-Einstein condensates which have been imprinted with a
topological configuration, in the manner of \cite{RuostekoskiA2001},
will relax via dissipation to the stable Skyrmion configurations we
describe here.  Figure 1 (c,d) shows the distribution of
component-$|2\rangle$ that would be observed experimentally for these
configurations.  Clearly, the essential condition for the formation of
stable Skyrmions is that there is {\it phase-separation} between the
two components. We therefore anticipate stable Skyrmions to exist for
small deviations from the limit $\Delta\ll \bar U$ studied. Phase
separation occurs provided\cite{HoS96} $U_{12}^2>U_{11} U_{22}$
(equivalent to $\Delta > 0$ for $\Delta \ll \bar U$).  The observation
of phase separation in the system $^{87}$Rb $|F=2,m_F=1\rangle$
$|1,-1\rangle$\cite{HallMEWC98}, makes this a candidate system in
which the
stable Skyrmions may be realised.

Our research is funded by EPSRC (PMS) and PPARC (RAB). The parallel
computations were performed at the National Cosmology Supercomputing
Centre in Cambridge.

\widetext 

\end{document}